%% file: tajima24.2.19.tex
\documentclass[a4paper,11pt]{article}
\pdfoutput=1
\usepackage{jinstpub}

\title{Antiproton beams with low energy spread for antihydrogen production}

\author[a,b,1]{M.~Tajima,\note{Corresponding author. Present address: RIKEN Nishina Center for Accelerator-Based Science.}}
\author[a]{N.~Kuroda,}
\author[c]{C.~Amsler,}
\author[b]{H.~Breuker,}
\author[d]{C.~Evans,}
\author[c,2]{M.~Fleck,\note{Present address: Institute of Physics, University of Tokyo}}
\author[c]{A.~Gligorova,}
\author[e]{H.~Higaki,}
\author[f]{Y.~Kanai,}
\author[c]{B.~Kolbinger,}
\author[c]{A.~Lanz,}
\author[d,g]{M.~Leali,}
\author[c,3]{V.~M\"{a}ckel,\note{Present address: Ulmer Fundamental Symmetries Laboratory, RIKEN.}}
\author[h]{C.~Malbrunot,}
\author[g,i]{V.~Mascagna,}
\author[a]{Y.~Matsuda,}
\author[c]{D.~Murtagh,}
\author[j]{Y.~Nagata,}
\author[c]{A.~Nanda,}
\author[b,4]{B.~Radics,\note{Present address: Institute for Particle Physics and Astrophysics, ETH Zurich, CH-8093 Zurich, Switzerland.}}
\author[c]{M.~Simon,}
\author[b]{S.~Ulmer,}
\author[d,g]{L.Venturelli,}
\author[c]{E.~Widmann,}
\author[c,5]{M.~Wiesinger,\note{Present address: Max Planck Institute for Nuclear Physics, D-69117 Heidelberg, Germany and Ulmer Fundamental Symmetries Laboratory, RIKEN.}}
\author[b]{and Y.~Yamazaki.}

\affiliation[a]{Institute of Physics, University of Tokyo,\\3-8-1 Komaba, Meguro, 153-8902 Tokyo, Japan}
\affiliation[b]{Ulmer Fundamental Symmetries Laboratory, RIKEN,\\2-1 Hirosawa, Wako, 351-0198 Saitama, Japan}
\affiliation[c]{Stefan Meyer Institute of the Austrian Academy of Sciences,\\Boltzmanngasse 3, 1090 Vienna, Austria}
\affiliation[d]{Department of Information Engineering, University of Brescia,\\Via Branze 38, 25123 Brescia, Italy}
\affiliation[e]{Graduate School of Advanced Sciences of Matter, Hiroshima University,\\1-3-2 Kagamiyama, Higashi-Hiroshima, 739-8530 Hiroshima, Japan}
\affiliation[f]{RIKEN Nishina Center for Accelerator-Based Science,\\2-1 Hirosawa, Wako, 351-0198 Saitama, Japan}
\affiliation[g]{National Institute for Nuclear Physics, Sezione di Pavia,\\Via Bassi 6, 27100 Pavia, Italy}
\affiliation[h]{Experimental Physics Department, CERN,\\1211 Geneva 23, Switzerland}
\affiliation[i]{Department of Science and High Technology, University of Insubria,\\Via Valleggio 11, 22100 Como, Italy}
\affiliation[j]{Department of Physics, Tokyo University of Science,\\1-3 Kagurazaka, Shinjuku, 162-8601 Tokyo, Japan}

\emailAdd{mtajima@riken.jp}

\abstract{A low energy antiproton transport from the ASACUSA's antiproton accumulation trap (MUSASHI trap) to the antihydrogen production trap (double cusp trap) is developed. The longitudinal antiproton energy spread after the transport line is $0.23\pm0.02$~eV, compared with $15$~eV with a previous method used in 2012. This reduction is achieved by an adiabatic transport beamline with several pulse-driven coaxial coils. Antihydrogen atoms are synthesized by directly injecting the antiprotons into a positron plasma, resulting in the higher production rate.}

\keywords{Beam optics, ion and atom traps, plasma diagnostics, antiproton, antihydrogen production.}

\begin{document}
\maketitle
\flushbottom

\section{Introduction}
Antihydrogen ($\overline{\rm H}$) with 11 candidates at 1.9 GeV/c was first reported at CERN's Low Energy Antiproton Ring (LEAR) in 1996~\cite{Baur1996}.
The first detection of cold $\overline{\rm H}$ by mixing trapped antiprotons ($\overline{\rm p}$s) with positrons ($e^+$s) was reported in 2002~\cite{Amoretti2002,Gabrielse2002} soon followed by the first magnetic trapping of $\overline{\rm H}$ in 2010~\cite{Andresen2010}.

Measuring atomic transitions in $\overline{\rm H}$ with high precision will be one of the most precise tests of the CPT symmetry between matter and antimatter.
A microwave spectroscopy experiment has recently measured the ground-state hyperfine splitting of trapped $\overline{\rm H}$ with a relative precision of $4\times10^{-4}$~\cite{Ahmadi2017}.
The ASACUSA collaboration intends to perform a Rabi-type experiment to measure the ground-state hyperfine splitting of $\overline{\rm H}$ using instead a spin-polarized $\overline{\rm H}$ beam~\cite{Kuroda2014}.
Measurements of the hyperfine splitting with a beam are less sensitive to magnetic field gradients to which hyperfine transitions are quite sensitive.

The ASACUSA $\overline{\rm H}$ beam is extracted from a cusped magnetic field which is generated by a pair of superconducting anti-Helmholtz coils~\cite{Kuroda2014,Mohri2003,Nagata2014}. 
A Rabi-type spectroscopy apparatus was designed to achieve a relative precision in the measurement of the ground-state hyperfine splitting of $\overline{\rm H}$ of less than a ppm.
The spectroscopy in a Rabi-type hydrogen beam was demonstrated and a relative precision of $2.7\times10^{-9}$ was achieved~\cite{Diermaier2017}.

\begin{figure}[h]
\centering
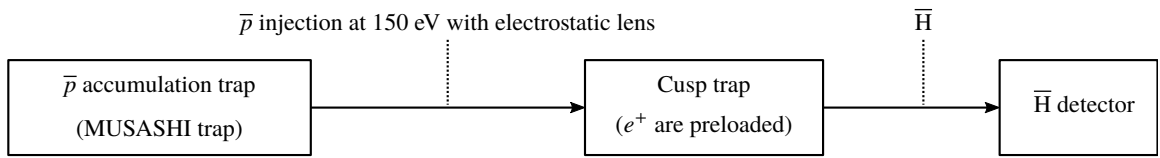
\caption{Diagram of the direct injection method used in 2012.}
\label{scheme2012}
\end{figure}
We have produced earlier $\overline{\rm H}$ atoms in the cusp trap (composed of Multi-ring electrodes~(MRE) in the cusp magnetic field) by directly injecting $\overline{p}$s into a pre-loaded $e^+$ plasma, following the so-called direct injection method~\cite{Enomoto2010} shown in figure~\ref{scheme2012}.
The $\overline{p}$s were transported towards the cusp trap at 150~eV with an electrostatic focusing lens system from the $\overline{p}$ accumulation trap (MUSASHI trap~\cite{Kuroda2012}).
With this method, the ASACUSA-Cusp collaboration succeeded in 2012 in detecting 80~$\overline{\rm H}$ atoms at the $\overline{\rm H}$ detector~\cite{Kuroda2014}.
Although it was the first demonstration of $\overline{\rm H}$ atoms detected downstream of their formation region, the rate was too small for the envisioned spectroscopy measurements initially with a precision of 1~ppm requiring about 4000~$\overline{\rm H}$.
One of the limiting factor was the longitudinal energy spread of the injected $\overline{p}$ which was measured to be $\sigma\sim15$~eV, assuming a Gaussian distribution~\cite{Minori2017}.
With the direct injection method, the kinetic energy of the injected $\overline{p}$s is adjusted to maximize the number of $\overline{p}$s in the cusp trap and to minimize the relative energy of $\overline{p}$ with respect to the pre-loaded $e^+$ plasma.
The relative longitudinal energy could not be smaller than 15~eV of the energy spread of injected $\overline{p}$, whereas the temperature of the $e^+$s was estimated to be about 0.02~eV (200~K)~\cite{Enomoto2010d}.
Hence the large relative energy caused heating of the $e^+$ plasma, which decreased the $\overline{\rm H}$ production rate because the rate strongly depends on the $e^+$ plasma temperature~\cite{Gabrielse1989}. 
With a low energy spread, a smaller relative energy would lead to a higher $\overline{\rm H}$ production rate.
Therefore a new $\overline{p}$ transport system was designed to reduce the longitudinal energy spread~\cite{Minori2017}. 
This paper describes a further development for $\overline{p}$ injection at 1.5~eV, which led to an observed energy distributions with energy spread $\sigma=0.23\pm0.02$~eV and resulted in the two orders of magnitude higher $\overline{\rm H}$ production rate.

\section{Experimental setup}
\begin{figure}[h]
\centering
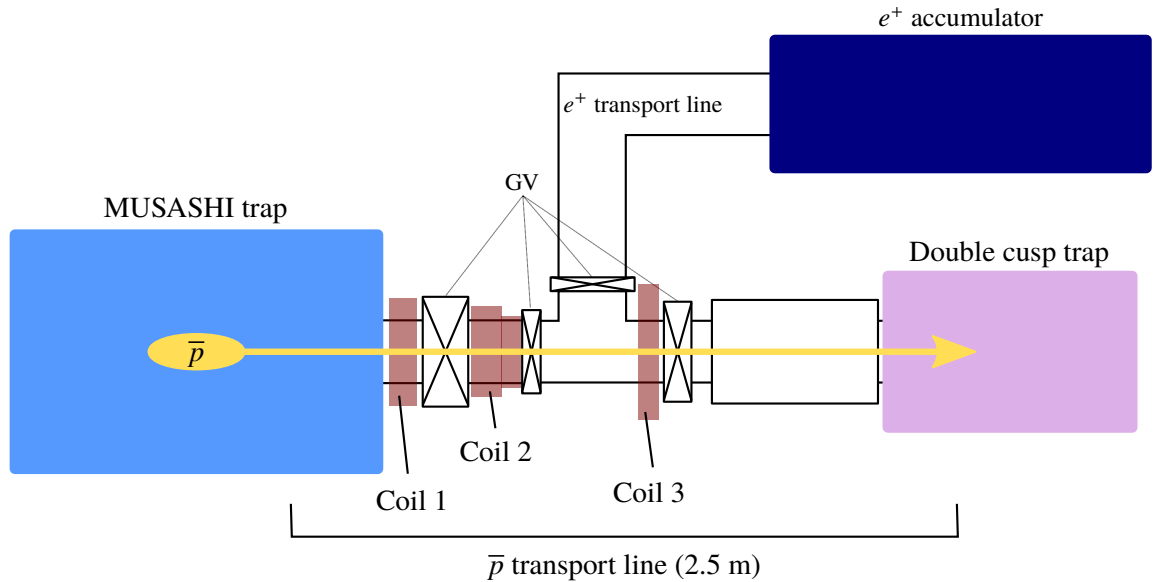
\caption{Sketch of the experimental setup (not to scale).}
\label{setup2016}
\end{figure}
Figure~\ref{setup2016} shows a sketch of the apparatus, including the MUSASHI trap, the $\overline{p}$ transport line, the $e^+$ accumulator, the $e^+$ transport line, and the $\overline{\rm H}$ production trap (double cusp trap).
The magnetic field of the double cusp trap is generated by two pairs of superconducting anti-Helmholtz coils.
Both MUSASHI trap and double cusp trap are of the Penning-Malmberg-type, which confines charged particles by static electromagnetic fields.
Electric fields are produced by the MRE for both traps.
The MRE are designed for the stable confinement of a non-neutral plasma~\cite{Mohri1998}.

A bunch of $3\times10^7$ $\overline{p}$s at 5.3~MeV is supplied every 2~min by the Antiproton Decelerator (AD) at CERN.
The $\overline{p}$s are decelerated to 100~keV by a radio frequency quadrupole decelerator~(RFQD)~\cite{Bylinsky2000}, and injected into the MUSASHI trap after further deceleration down to 10~keV by a thin foil.
They are first cooled through collisions with pre-loaded electrons ($\sim10^8$), and then radially compressed by applying a rotating-wall electric field (typically 247~kHz, 2~V peak to peak during 120~s) with fewer($\sim10^5$) electrons~\cite{Kuroda2008}.
A cloud of $2\times10^6$ $\overline{p}$s per 3~AD shots is typically prepared in the MUSASHI trap.

\begin{figure}[h]
\centering
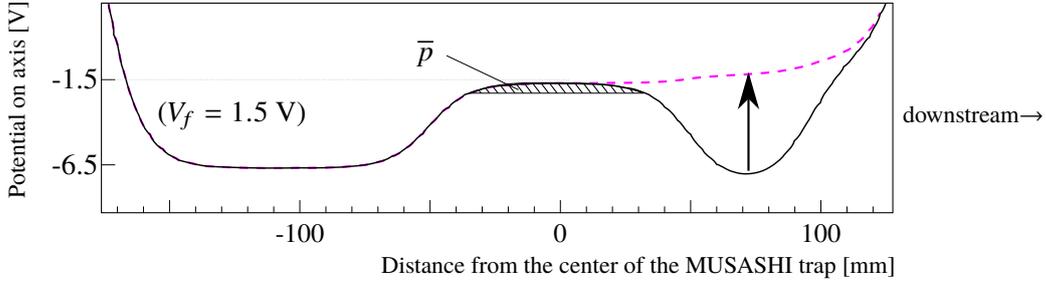
\caption{Electrostatic potential manipulations on the MUSASHI trap axis for $\overline{p}$ extraction. The solid and dashed lines show the configurations before and during $\overline{p}$ extraction. The hatched region corresponds to the $\overline{p}$ cloud. A pulsed voltage (rise time 50~ns, amplitude 5~V) is applied (arrow) to remove the potential barrier on the downstream side.}
\label{potmsh}
\end{figure}

Figure~\ref{potmsh} shows the electric field manipulations to extract the $\overline{p}$ cloud from the MUSASHI trap towards the $\overline{p}$ transport line. 
The MUSASHI MRE are floated at $-1.5$~V (hereinafter referred to as $V_f=1.5$~V), which corresponds to an ultra-slow beam extraction energy of 1.5~eV. 
The solid line shows the electrostatic potential configuration before extraction.
A pulsed voltage is applied to remove the potential barrier at the downstream side. 
The dashed line shows the potential configuration at the moment of extraction.
It is designed to minimize the potential gradient in the $\overline{p}$ region, and optimized experimentally to maximize the trapping efficiency in the double cusp trap.

\begin{figure}[h]
\centering
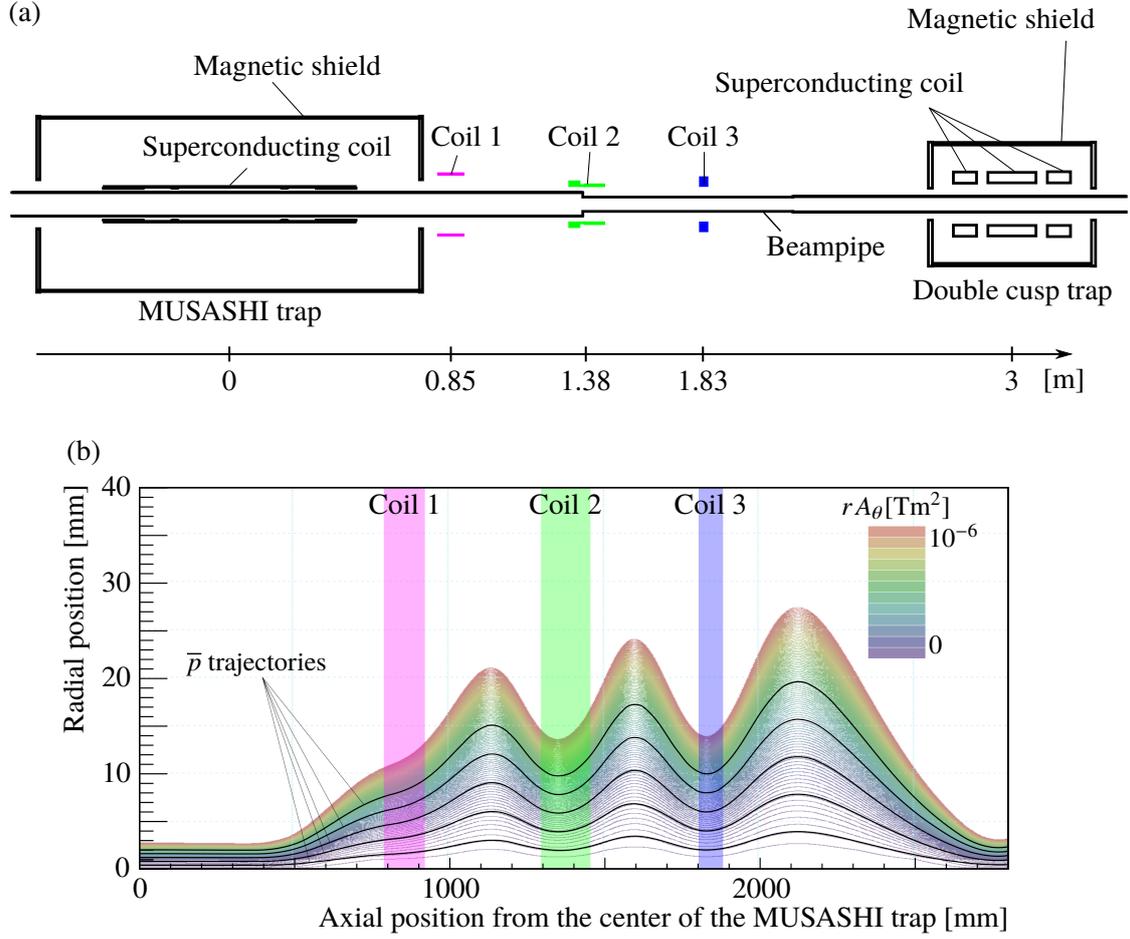
\caption{(a) Schematic view of the $\overline{p}$ transport line. (b) Simulation results for $\overline{p}$ trajectories along the transport line (black solid curves). The initial radial positions are 0.4, 0.8, 1.2, 1.6, and 2.0~mm. The initial kinetic energy of the $\overline{p}$s is 1.5~eV. The colored stripes correspond to the locations of the three transport coils. The extraction process in the MUSASHI trap is not included in this simulations. The magnetic field lines lie along the contour lines of $rA_\theta$ where $A_\theta$ is a non-zero component of the vector potential.}
\label{pbartraj}
\end{figure}

\begin{figure}[h]
\centering
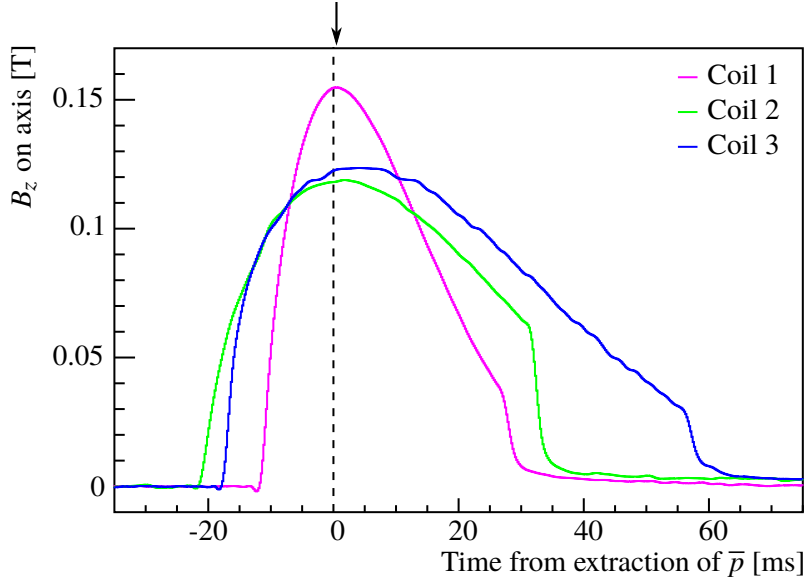
\caption{Axial magnetic field strength $B_z$ on axis of the three transport coils as a function of $\overline{p}$ extraction time. $\overline{p}$s reach the double cusp at 150~$\mu$s (arrow).}
\label{coilcurrent}
\end{figure}

The $\overline{p}$ beamline with three transport coils is designed so that $\overline{p}$s adiabatically follow the magnetic field lines.
Figure~\ref{pbartraj}~(a) shows a sketch of the setup along the transport line, and figure~\ref{pbartraj}~(b) shows simulation results of the $\overline{p}$ trajectories along the transport line. 
Commercially available software (Tricomp by Field Precision LLC) is used for the simulation. 
Firstly, the cylindrical electrostatic and magnetostatic fields is calculated using a finite element method and then the $\overline{p}$ trajectory in the static fields is simulated.
The initial kinetic energy of each $\overline{p}$ is set to 1.5~eV.
The trajectories with initial radial positions of 0.4, 0.8, 1.2, 1.6, and 2.0~mm are shown.
Coil~1 is located next to the MUSASHI trap to suppress the radial expansion of the $\overline{p}$ cloud due to the drastic divergence of the magnetic field lines at the exit of the MUSASHI trap.
In order to achieve the required magnetic field strength for the adiabatic transport, large currents have to be applied to the three transport coils.
Therefore they are energized by pulsed currents with time widths of about 50~ms to suppress Joule heating.
Table~\ref{coils} shows specifications of the three coils.
Figure~\ref{coilcurrent} shows the axial magnetic field $B_z$ on the axis of each transport coil as a function of time, estimated from the monitored currents when the coils are energized.
Since the time of flight of the $\overline{p}$s along the transport line is of the order of $\mu$s, the magnetic field can be considered constant during $\overline{p}$ transport.

\begin{table}[h]
\centering
\begin{tabular}{|c|c|c|c|c|c|c|c|}\hline
             &   {\small Inner}&   {\small Outer}& {\small Axial}&              &                   &                   &               Maximum\\
             &{\small diameter}&{\small diameter}&{\small length}&              &{\small Resistance}&{\small Inductance}&{\small $B_z$ on axis}\\
{\small Name}&             [mm]&             [mm]&           [mm]&{\small Turns}&         [$\Omega$]&               [mH]&                   [T]\\ \hline
       Coil~1&              400&              408&             96&            48&               0.13&               1.35&                  0.15\\ \hline
       Coil~2&              250&              306&            115&           150&               0.41&               5.06&                  0.12\\ \hline
       Coil~3&              250&              360&             30&           147&               0.44&               10.2&                  0.12\\ \hline
\end{tabular}
\caption{Parameters of the three transport coils.}
\label{coils}
\end{table}

\section{Antiproton energy distribution}
\begin{figure}[h]
\centering
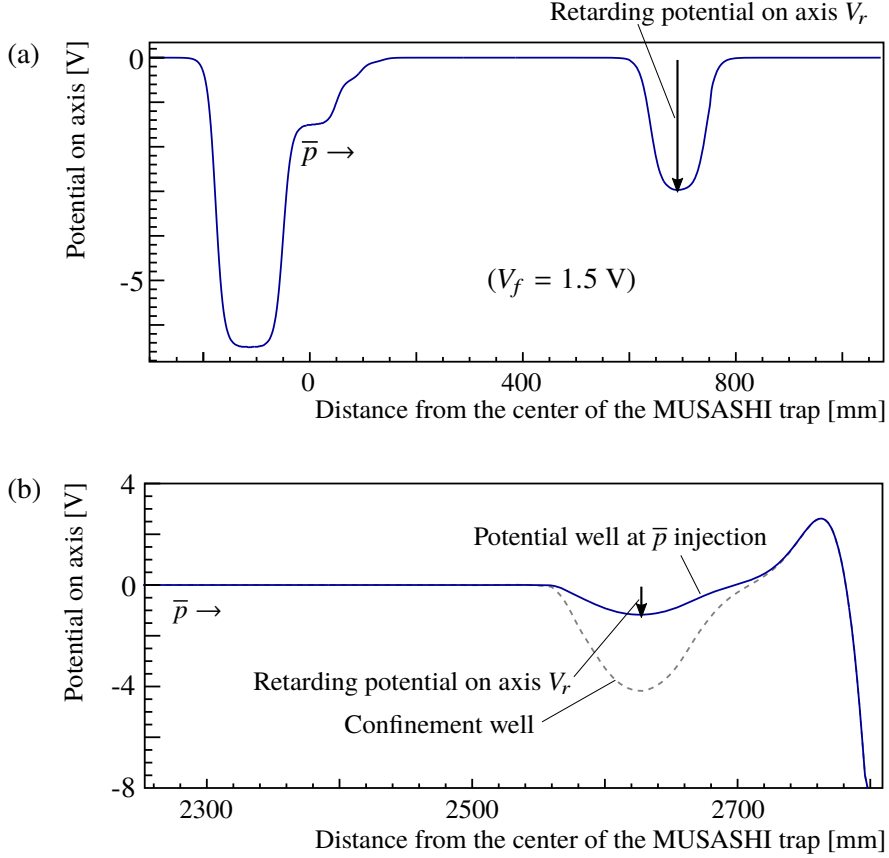
\caption{Electrostatic potentials on axis for the measurement of the energy distribution of $\overline{p}$s at the MUSASHI trap (a) and the double cusp trap (b).}
\label{retardpot}
\end{figure}

The energy distribution of the $\overline{p}$s is measured by using retarding potentials at the exit of the MUSASHI trap and at the entrance of the double cusp trap.
Figure~\ref{retardpot}~(a) shows the potential distribution on the axis at the exit of the MUSASHI trap.
Additional cylindrical electrodes separated from the MUSASHI MRE, located 650~mm downstream of the center of the MUSASHI trap, are used to generate the retarding potential barrier. 
A gate valve located 1.9~m downstream is closed during the measurement.
The $\overline{p}$s are extracted from the MUSASHI MRE towards the potential barrier and the transmitted particles annihilate on the surface of the closed gate valve.
Charged annihilation particles (mainly 3 charged pions on average) are detected as a pulsed current by a 2~cm-thick plastic scintillator equipped with a photomultiplier (H7195).
When the $\overline{p}$s are released, the longitudinal energy distribution is given by the detected charge, which is propotional to the number of extracted $\overline{p}$s (within 10~\%) as a function of the retarding potential on the beam axis ($V_r$).

Figure~\ref{retardpot}~(b) shows the potential distribution for the measurement of the energy distribution at the entrance of the double cusp trap.
Firstly, an electrostatic potential well is prepared in the double cusp trap. 
The $\overline{p}$s are then transported from the MUSASHI trap and injected into the double cusp trap.
The solid line shows the configuration at $\overline{p}$ injection.
The upstream side of the potential well is closed (dotted line) after a given time (hereinafter referred to as catching time), which has been optimized to trap as many $\overline{p}$s as possible.
Finally, the trapped $\overline{p}$s in the potential well are slowly released to annihilate on the surface of surrounding materials.
The annihilations are detected by 8 plastic scintillator bars surrounding the vacuum pipe of the double cusp trap~\cite{Radics2015}.

\begin{figure}[h]
\centering
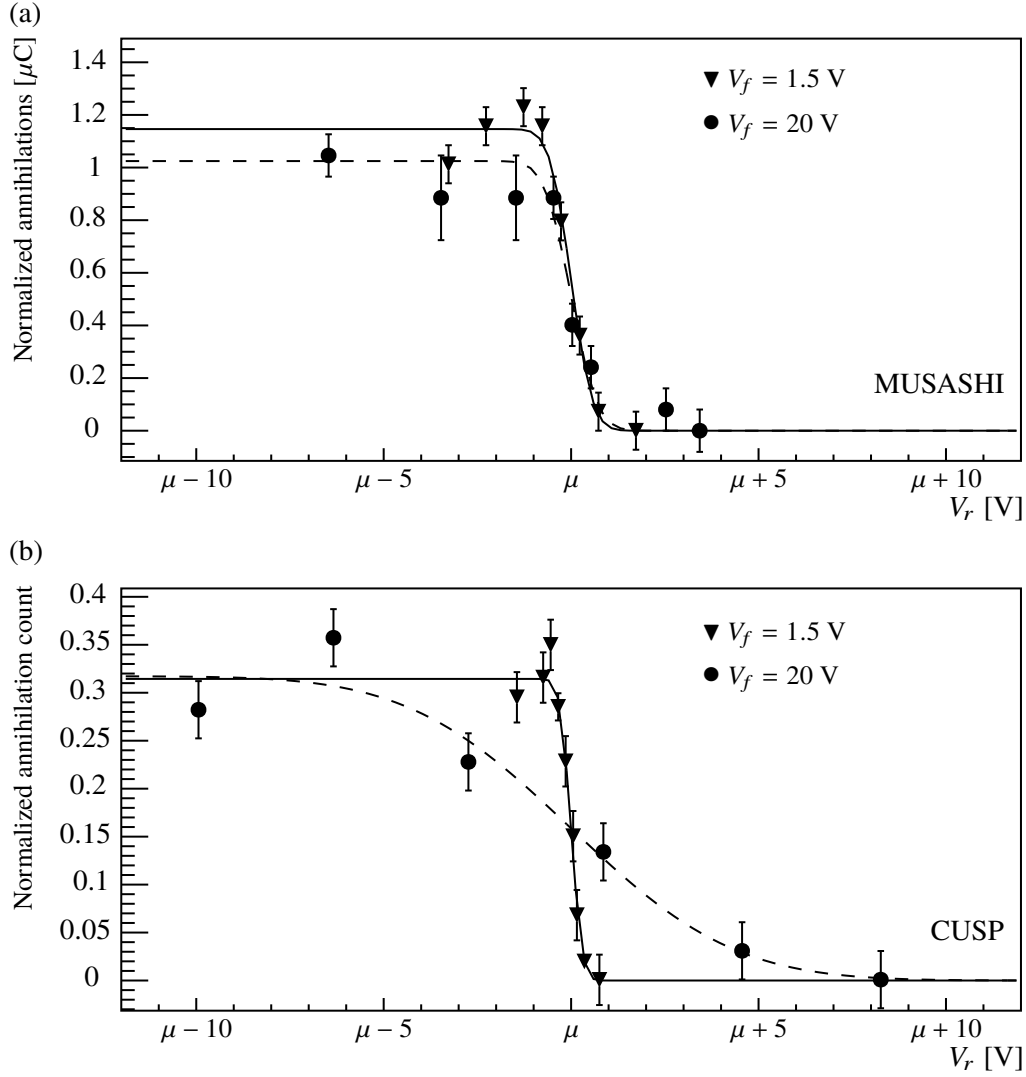
	\caption{Longitudinal energy distributions of the $\overline{p}$s at the exit of the MUSASHI trap (a) and at the entrance of the double cusp trap (b), for two different float voltages $V_f$. The curves are fits using \eqref{eqfit}.}
\label{edist}
\end{figure}

The measurements are repeated with different retarding potentials. 
At each trial, the number of confined $\overline{p}$s in the MUSASHI trap is monitored separately on the upstream side, to which the annihilation signals are normalized.
Figure~\ref{edist}~(a) and (b) show the results obtained at the MUSASHI trap exit and at the double cusp trap entrance, together with fits $F(V_r)$ given by 
\begin{equation}
\label{eqfit}
F(V_r)=\frac{C}{2}\left(1+{\rm erf}\left[\frac{V_r-\mu}{\sqrt{2\sigma^2}}\right]\right),
\end{equation}
which is a cumulative distribution function of a Gaussian distribution with mean $\mu$, standard deviation $\sigma$, and normalization constant $C$.
We also show the distribution for $V_f=20$~V, comparable to the one reported in~\cite{Minori2017}. 
Table~\ref{spread} shows the standard deviation $\sigma$, a measure of the longitudinal energy spread. 

The longitudinal energy spread at the exit of the MUSASHI trap is consistent for both $V_f=1.5$~V and $V_f=20$~V.
At the entrance of the double cusp trap, the longitudinal energy spread is broader for $V_f=20$~V.
This can be qualitatively explained by the fact that the transport condition deviates from adiabatic condition as $\overline{p}$s travel faster.
In other words, $\overline{p}$s are not slow enough compared to the change of magnetic field strength along their trajectories and the longitudinal energy spread does not conserve. 
With $V_f=1.5$~V, the energy spread of $0.23\pm0.02$~eV is about two orders of magnitude smaller than 15~eV, obtained by using our previous transport scheme in 2012~\cite{Minori2017}.
The energy spread at the double cusp trap entrance for $V_f=1.5$~V is smaller than the value at the MUSASHI trap exit, which can be explained by the loss of fast particles during catching, mainly those which have a shorter travelling time in the double cusp trap. 
The number of trapped $\overline{p}$s per 3~AD shots is typically $6\times10^5$ out of $2\times10^6$ confined in the MUSASHI trap.

\begin{table}[h]
\centering
\begin{tabular}{|c|c|c|}\hline
      Initial & \multicolumn{2}{c|}{Energy spread $\sigma$ [eV]}\\ \cline{2-3}
float voltage &          at the exit &  at the entrance of  \\
    $V_f$ [V] &  of the MUSASHI trap & the double cusp trap \\ \hline
          1.5 &        $0.45\pm0.08$ &       $0.23\pm0.02$  \\ \hline
           20 &         $0.6\pm0.1$  &        $3.4 \pm0.8$  \\ \hline
\end{tabular}
\caption{Longitudinal energy spread expressed as the standard deviation $\sigma$ in \eqref{eqfit}.}
\label{spread}
\end{table}

\section{$\overline{\rm H}$ production rate}
\begin{figure}[h]
\centering
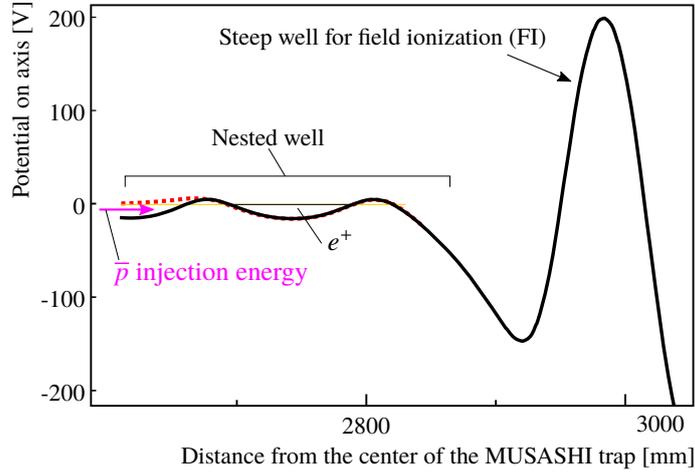
\caption{Electrostatic potential on axis inside the double cusp trap for $\overline{\rm H}$ production. The nested well at the upstream side confines both $e^+$s and $\overline{p}$s. The steep well at the downstream side is for field ionizing $\overline{\rm H}$ atoms. The dotted and solid lines show the configurations before and after $\overline{p}$ injection, respectively.}
\label{potfi}
\end{figure}

$\overline{\rm H}$ atoms are produced with the direct injection method by using the $\overline{p}$ beam with reduced energy spread of $0.23\pm0.02$~eV.
Typically, a cloud of $6.0\times 10^5$ $\overline{p}$s with a radius of 2~mm are injected into a plasma of $1.2\times 10^8$ $e^+$s with the density of $6.0\times 10^8$cm$^{-3}$ and a radius of 1~mm. 
The number of $\overline{\rm H}$ atoms formed is estimated using the field ionization (FI) technique~\cite{Gabrielse2002} inside the double cusp trap.
Figure~\ref{potfi} shows the electrostatic potential on axis for $\overline{\rm H}$ production.
The upstream nested well confines the positively charged pre-loaded $e^+$ plasma at the center and the negatively charged $\overline{p}$ cloud.
A steep potential well (FI) with a strong electrostatic field is used to field-ionize the $\overline{\rm H}$ atoms in the double cusp trap.
Rydberg $\overline{\rm H}$ atoms ($n \ge 40$) reaching the FI~well are ionized by the strong field thus confining the remaining $\overline{p}$ inside the FI~well.
When the confined $\overline{p}$s in the FI~well are released, they annihilate on the surrounding materials and the emitted charged annihilation particles 
are detected by plastic scintillators.
The solid angle of the FI~well from the center of the nested well is around 1~\%.
The time evolution of $\overline{\rm H}$ production could be studied by periodically releasing the trapped $\overline{p}$s in the FI~well~\cite{Enomoto2010}.

\begin{figure}[h]
\centering
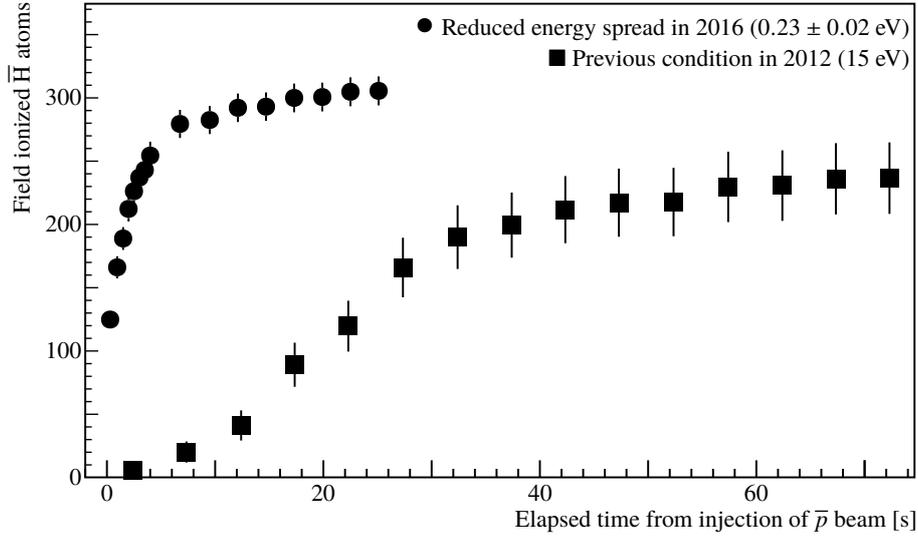
\caption{Time evolution of the number of field ionized $\overline{\rm H}$ atoms for two different conditions (cumulative plot). The number of $\overline{p}$s confined in the nested well for $\overline{\rm H}$ production is $3\times 10^5$ for both cases.}
\label{hbarint}
\end{figure}

Figure~\ref{hbarint} shows the observed time evolution for $\overline{\rm H}$ production. 
The squares show the previous result in 2012 as reference~\cite{Kuroda2014}. 
The average number of field ionized $\overline{\rm H}$ atoms is 6 during the first 5~s of mixing.
The circles show the result using the $\overline{p}$s with reduced energy spread.
The number of field ionized $\overline{\rm H}$ atoms increases to 260 during the first 4~s.
In fact, the highest rate reached already $208\pm7$ s$^{-1}$ during the first 0.7~s.
This indicates a weaker heating of $e^+$s by the injected $\overline{p}$s with reduced energy spread.
The higher production rate increases the signal/background ratio at the downstream $\overline{\rm H}$ detector (background events at the detector are essentially due to cosmic rays~\cite{Nagata2018}). 
Note that it is unlikely that the increase in FI atoms is due to the production of higher Rydberg states, since the lower the $e^+$ temperature, the lower the produced Rydberg states~\cite{Radics2014}.
The principal quantum number $n$ distribution is measured with the $\overline{\rm H}$ atoms extracted downstream of the double cusp trap. 
Among 43~mixing runs, 7~$\overline{\rm H}$ events with $n<14$ are recorded, as declared in~\cite{Malbrunot2018}. 
If the on-axis axial velocity is small (less than roughly 1300 m${\rm s}^{-1}$), those $\overline{\rm H}$ atoms would have been in ground-state when reaching the detector. 

\section{Conclusion}
A low energy $\overline{p}$ transport method from the MUSASHI trap to the double cusp trap is developed.
With the adiabatic transport beamline containing several pulse-driven coaxial coils, the longitudinal energy spread of $\overline{p}$s is reduced to $0.23\pm0.02$~eV, compared with 15~eV obtained earlier.
This results in a two orders of magnitude higher $\overline{\rm H}$ production rate. 

\acknowledgments
This work was supported by the Grant-in-Aid for Specially Promoted Research 24000008 of Japanese MEXT, Special Research Projects for Basic Science of RIKEN, European Research Council under European Union's Seventh Framework Programme (FP7/2007-2013) /ERC Grant Agreement (291242), the Austrian Ministry of Science and Research, Austrian Science Fund (FWF): W1252-N27, and Universit\`a di Brescia and Istituto Nazionale di Fisica Nucleare.
MT was supported by the Junior Research Associate program of RIKEN.

\end{document}

%% file: scheme2012.pdf_tex
\begingroup%
  \makeatletter%
  \providecommand\color[2][]{%
    \errmessage{(Inkscape) Color is used for the text in Inkscape, but the package 'color.sty' is not loaded}%
    \renewcommand\color[2][]{}%
  }%
  \providecommand\transparent[1]{%
    \errmessage{(Inkscape) Transparency is used (non-zero) for the text in Inkscape, but the package 'transparent.sty' is not loaded}%
    \renewcommand\transparent[1]{}%
  }%
  \providecommand\rotatebox[2]{#2}%
  \ifx\svgwidth\undefined%
    \setlength{\unitlength}{429.86595334bp}%
    \ifx\svgscale\undefined%
      \relax%
    \else%
      \setlength{\unitlength}{\unitlength * \real{\svgscale}}%
    \fi%
  \else%
    \setlength{\unitlength}{\svgwidth}%
  \fi%
  \global\let\svgwidth\undefined%
  \global\let\svgscale\undefined%
  \makeatother%
  \begin{picture}(1,0.15107608)%
    \put(0,0){\includegraphics[width=\unitlength,page=1]{scheme2012.pdf}}%
    \put(0.38373865,0.11037774){\color[rgb]{0,0,0}\makebox(0,0)[b]{\smash{{\footnotesize $\overline{p}$ injection at 150 eV with electrostatic lens}}}}%
    \put(0.7964632,0.11009604){\color[rgb]{0,0,0}\makebox(0,0)[b]{\smash{{\footnotesize $\overline{\rm H}$}}}}%
    \put(0.13372594,0.05638943){\color[rgb]{0,0,0}\makebox(0,0)[b]{\smash{{\footnotesize $\overline{p}$ accumulation trap}}}}%
    \put(0.60613736,0.05638943){\color[rgb]{0,0,0}\makebox(0,0)[b]{\smash{{\footnotesize Cusp trap}}}}%
    \put(0.9361331,0.03874289){\color[rgb]{0,0,0}\makebox(0,0)[b]{\smash{{\footnotesize $\overline{\rm H}$ detector}}}}%
    \put(0.60613736,0.02127804){\color[rgb]{0,0,0}\makebox(0,0)[b]{\smash{{\footnotesize ($e^+$ are preloaded)}}}}%
    \put(0.13372593,0.01788326){\color[rgb]{0,0,0}\makebox(0,0)[b]{\smash{{\footnotesize (MUSASHI trap)}}}}%
  \end{picture}%
\endgroup%

%% file: setup2016.pdf_tex
\begingroup%
  \makeatletter%
  \providecommand\color[2][]{%
    \errmessage{(Inkscape) Color is used for the text in Inkscape, but the package 'color.sty' is not loaded}%
    \renewcommand\color[2][]{}%
  }%
  \providecommand\transparent[1]{%
    \errmessage{(Inkscape) Transparency is used (non-zero) for the text in Inkscape, but the package 'transparent.sty' is not loaded}%
    \renewcommand\transparent[1]{}%
  }%
  \providecommand\rotatebox[2]{#2}%
  \ifx\svgwidth\undefined%
    \setlength{\unitlength}{437.30702798bp}%
    \ifx\svgscale\undefined%
      \relax%
    \else%
      \setlength{\unitlength}{\unitlength * \real{\svgscale}}%
    \fi%
  \else%
    \setlength{\unitlength}{\svgwidth}%
  \fi%
  \global\let\svgwidth\undefined%
  \global\let\svgscale\undefined%
  \makeatother%
  \begin{picture}(1,0.50689623)%
    \put(0,0){\includegraphics[width=\unitlength,page=1]{setup2016.pdf}}%
    \put(0.82402472,0.49023157){\color[rgb]{0,0,0}\makebox(0,0)[b]{\smash{{\small $e^+$ accumulator}}}}%
    \put(0.17159051,0.32641957){\color[rgb]{0,0,0}\makebox(0,0)[b]{\smash{MUSASHI trap}}}%
    \put(0.86548192,0.28820624){\color[rgb]{0,0,0}\makebox(0,0)[b]{\smash{Double cusp trap}}}%
    \put(0.5374685,0.01927606){\color[rgb]{0,0,0}\makebox(0,0)[b]{\smash{$\overline{p}$ transport line (2.5~m)}}}%
    \put(0.45004041,0.34929131){\color[rgb]{0,0,0}\makebox(0,0)[b]{\smash{{\footnotesize GV}}}}%
    \put(0.49259516,0.42761339){\color[rgb]{0,0,0}\makebox(0,0)[lb]{\smash{}}}%
    \put(0.48506715,0.41689712){\color[rgb]{0,0,0}\makebox(0,0)[lb]{\smash{{\footnotesize $e^+$ transport line}}}}%
    \put(0.35637396,0.07456828){\color[rgb]{0,0,0}\makebox(0,0)[b]{\smash{Coil~1}}}%
    \put(0.42776405,0.11648913){\color[rgb]{0,0,0}\makebox(0,0)[b]{\smash{Coil~2}}}%
    \put(0.55855271,0.08188576){\color[rgb]{0,0,0}\makebox(0,0)[b]{\smash{Coil~3}}}%
    \put(0.17222287,0.2029439){\color[rgb]{0,0,0}\makebox(0,0)[b]{\smash{$\overline{p}$}}}%
  \end{picture}%
\endgroup%

%% file: potmsh.pdf_tex
\begingroup%
  \makeatletter%
  \providecommand\color[2][]{%
    \errmessage{(Inkscape) Color is used for the text in Inkscape, but the package 'color.sty' is not loaded}%
    \renewcommand\color[2][]{}%
  }%
  \providecommand\transparent[1]{%
    \errmessage{(Inkscape) Transparency is used (non-zero) for the text in Inkscape, but the package 'transparent.sty' is not loaded}%
    \renewcommand\transparent[1]{}%
  }%
  \providecommand\rotatebox[2]{#2}%
  \ifx\svgwidth\undefined%
    \setlength{\unitlength}{428.77376526bp}%
    \ifx\svgscale\undefined%
      \relax%
    \else%
      \setlength{\unitlength}{\unitlength * \real{\svgscale}}%
    \fi%
  \else%
    \setlength{\unitlength}{\svgwidth}%
  \fi%
  \global\let\svgwidth\undefined%
  \global\let\svgscale\undefined%
  \makeatother%
  \begin{picture}(1,0.24917435)%
    \put(0,0){\includegraphics[width=\unitlength,page=1]{potmsh.pdf}}%
    \put(0.05000169,0.25103339){\color[rgb]{0,0,0}\rotatebox{90}{\makebox(0,0)[rb]{\smash{{\footnotesize Potential on axis [V]}}}}}%
    \put(0.74240973,0.03973701){\color[rgb]{0,0,0}\makebox(0,0)[b]{\smash{{\small 100}}}}%
    \put(0.40536178,0.19867562){\color[rgb]{0,0,0}\makebox(0,0)[rb]{\smash{$\overline{p}$}}}%
    \put(0.11365026,0.17378815){\color[rgb]{0,0,0}\makebox(0,0)[rb]{\smash{-1.5}}}%
    \put(0.11365026,0.09900231){\color[rgb]{0,0,0}\makebox(0,0)[rb]{\smash{-6.5}}}%
    \put(0.28912382,0.03973701){\color[rgb]{0,0,0}\makebox(0,0)[b]{\smash{{\small -100}}}}%
    \put(0.51576684,0.03973701){\color[rgb]{0,0,0}\makebox(0,0)[b]{\smash{{\small 0}}}}%
    \put(0.80738737,0.0122831){\color[rgb]{0,0,0}\makebox(0,0)[rb]{\smash{{\footnotesize Distance from the center of the MUSASHI trap [mm]}}}}%
    \put(0.81378319,0.14436124){\color[rgb]{0,0,0}\makebox(0,0)[lb]{\smash{{\footnotesize downstream$\rightarrow$}}}}%
    \put(0.16529674,0.14513334){\color[rgb]{0,0,0}\makebox(0,0)[lb]{\smash{($V_f=1.5$~V)}}}%
  \end{picture}%
\endgroup%

%% file: pbartraj.pdf_tex
\begingroup%
  \makeatletter%
  \providecommand\color[2][]{%
    \errmessage{(Inkscape) Color is used for the text in Inkscape, but the package 'color.sty' is not loaded}%
    \renewcommand\color[2][]{}%
  }%
  \providecommand\transparent[1]{%
    \errmessage{(Inkscape) Transparency is used (non-zero) for the text in Inkscape, but the package 'transparent.sty' is not loaded}%
    \renewcommand\transparent[1]{}%
  }%
  \providecommand\rotatebox[2]{#2}%
  \newcommand*\fsize{\dimexpr\f@size pt\relax}%
  \newcommand*\lineheight[1]{\fontsize{\fsize}{#1\fsize}\selectfont}%
  \ifx\svgwidth\undefined%
    \setlength{\unitlength}{416.66471226bp}%
    \ifx\svgscale\undefined%
      \relax%
    \else%
      \setlength{\unitlength}{\unitlength * \real{\svgscale}}%
    \fi%
  \else%
    \setlength{\unitlength}{\svgwidth}%
  \fi%
  \global\let\svgwidth\undefined%
  \global\let\svgscale\undefined%
  \makeatother%
  \begin{picture}(1,0.83562146)%
    \lineheight{1}%
    \setlength\tabcolsep{0pt}%
    \put(0,0){\includegraphics[width=\unitlength,page=1]{pbartraj.pdf}}%
    \put(0.89711902,0.5700408){\color[rgb]{0,0,0}\makebox(0,0)[t]{\lineheight{0}\smash{\begin{tabular}[t]{c}Double cusp trap\end{tabular}}}}%
    \put(0.67693016,0.61043517){\color[rgb]{0,0,0}\makebox(0,0)[lt]{\lineheight{0}\smash{\begin{tabular}[t]{l}Beampipe\end{tabular}}}}%
    \put(0.48122995,0.70888608){\color[rgb]{0,0,0}\makebox(0,0)[lt]{\lineheight{0}\smash{\begin{tabular}[t]{l}Coil~2\end{tabular}}}}%
    \put(0.58822482,0.70888608){\color[rgb]{0,0,0}\makebox(0,0)[lt]{\lineheight{0}\smash{\begin{tabular}[t]{l}Coil~3\end{tabular}}}}%
    \put(0.37664367,0.70888608){\color[rgb]{0,0,0}\makebox(0,0)[lt]{\lineheight{0}\smash{\begin{tabular}[t]{l}Coil~1\end{tabular}}}}%
    \put(0.10958746,0.03115497){\color[rgb]{0,0,0}\makebox(0,0)[lt]{\lineheight{0}\smash{\begin{tabular}[t]{l}0\end{tabular}}}}%
    \put(0.36784219,0.03115497){\color[rgb]{0,0,0}\makebox(0,0)[lt]{\lineheight{0}\smash{\begin{tabular}[t]{l}1000\end{tabular}}}}%
    \put(0.6460688,0.03115497){\color[rgb]{0,0,0}\makebox(0,0)[lt]{\lineheight{0}\smash{\begin{tabular}[t]{l}2000\end{tabular}}}}%
    \put(0.09392436,0.04960168){\color[rgb]{0,0,0}\makebox(0,0)[lt]{\lineheight{0}\smash{\begin{tabular}[t]{l}0\end{tabular}}}}%
    \put(0.08168228,0.13542249){\color[rgb]{0,0,0}\makebox(0,0)[lt]{\lineheight{0}\smash{\begin{tabular}[t]{l}10\end{tabular}}}}%
    \put(0.08168207,0.22124319){\color[rgb]{0,0,0}\makebox(0,0)[lt]{\lineheight{0}\smash{\begin{tabular}[t]{l}20\end{tabular}}}}%
    \put(0.08169787,0.30707477){\color[rgb]{0,0,0}\makebox(0,0)[lt]{\lineheight{0}\smash{\begin{tabular}[t]{l}30\end{tabular}}}}%
    \put(0.08168228,0.39288481){\color[rgb]{0,0,0}\makebox(0,0)[lt]{\lineheight{0}\smash{\begin{tabular}[t]{l}40\end{tabular}}}}%
    \put(0.89373825,0.00426815){\color[rgb]{0,0,0}\makebox(0,0)[rt]{\lineheight{0}\smash{\begin{tabular}[t]{r}Axial position from the center of the MUSASHI trap [mm]\end{tabular}}}}%
    \put(0.06519186,0.40355241){\color[rgb]{0,0,0}\rotatebox{90}{\makebox(0,0)[rt]{\lineheight{0}\smash{\begin{tabular}[t]{r}Radial position [mm]\end{tabular}}}}}%
    \put(0.82211167,0.25079174){\color[rgb]{0,0,0}\makebox(0,0)[lt]{\lineheight{0}\smash{\begin{tabular}[t]{l}{\small 0}\end{tabular}}}}%
    \put(0.82191018,0.34779917){\color[rgb]{0,0,0}\makebox(0,0)[lt]{\lineheight{0}\smash{\begin{tabular}[t]{l}{\small $10^{-6}$}\end{tabular}}}}%
    \put(0.7929325,0.37630586){\color[rgb]{0,0,0}\makebox(0,0)[t]{\lineheight{0}\smash{\begin{tabular}[t]{c}{\small $rA_{\theta}$[Tm$^2$]}\end{tabular}}}}%
    \put(0.19660513,0.55159427){\color[rgb]{0,0,0}\makebox(0,0)[t]{\lineheight{0}\smash{\begin{tabular}[t]{c}MUSASHI trap\end{tabular}}}}%
    \put(0.05029019,0.42563504){\color[rgb]{0,0,0}\makebox(0,0)[lt]{\lineheight{0}\smash{\begin{tabular}[t]{l}(b)\end{tabular}}}}%
    \put(-0.00149818,0.82183708){\color[rgb]{0,0,0}\makebox(0,0)[lt]{\lineheight{0}\smash{\begin{tabular}[t]{l}(a)\end{tabular}}}}%
    \put(0.35317365,0.37694823){\color[rgb]{0,0,0}\makebox(0,0)[t]{\lineheight{0}\smash{\begin{tabular}[t]{c}Coil~1\end{tabular}}}}%
    \put(0.49692384,0.37694823){\color[rgb]{0,0,0}\makebox(0,0)[t]{\lineheight{0}\smash{\begin{tabular}[t]{c}Coil~2\end{tabular}}}}%
    \put(0.62682064,0.37694823){\color[rgb]{0,0,0}\makebox(0,0)[t]{\lineheight{0}\smash{\begin{tabular}[t]{c}Coil~3\end{tabular}}}}%
    \put(0.15803446,0.23553397){\color[rgb]{0,0,0}\makebox(0,0)[lt]{\lineheight{0}\smash{\begin{tabular}[t]{l}{\small $\overline{p}$ trajectories}\end{tabular}}}}%
    \put(0.3413492,0.70010271){\color[rgb]{0,0,0}\makebox(0,0)[rt]{\lineheight{0}\smash{\begin{tabular}[t]{r}Superconducting coil\end{tabular}}}}%
    \put(0.85380499,0.75767492){\color[rgb]{0,0,0}\makebox(0,0)[rt]{\lineheight{0}\smash{\begin{tabular}[t]{r}Superconducting coil\end{tabular}}}}%
    \put(0.16386119,0.77301495){\color[rgb]{0,0,0}\makebox(0,0)[lt]{\lineheight{0}\smash{\begin{tabular}[t]{l}Magnetic shield\end{tabular}}}}%
    \put(0.97079103,0.8153821){\color[rgb]{0,0,0}\makebox(0,0)[rt]{\lineheight{0}\smash{\begin{tabular}[t]{r}Magnetic shield\end{tabular}}}}%
    \put(0.19600778,0.48898601){\color[rgb]{0,0,0}\makebox(0,0)[t]{\lineheight{0}\smash{\begin{tabular}[t]{c}0\end{tabular}}}}%
    \put(0.89643826,0.48898052){\color[rgb]{0,0,0}\makebox(0,0)[t]{\lineheight{0}\smash{\begin{tabular}[t]{c}3\end{tabular}}}}%
    \put(0.94305743,0.4907658){\color[rgb]{0,0,0}\makebox(0,0)[t]{\lineheight{0}\smash{\begin{tabular}[t]{c}[m]\end{tabular}}}}%
    \put(0.39425678,0.48898601){\color[rgb]{0,0,0}\makebox(0,0)[t]{\lineheight{0}\smash{\begin{tabular}[t]{c}0.85\end{tabular}}}}%
    \put(0.51504455,0.48898601){\color[rgb]{0,0,0}\makebox(0,0)[t]{\lineheight{0}\smash{\begin{tabular}[t]{c}1.38\end{tabular}}}}%
    \put(0.61988889,0.48898601){\color[rgb]{0,0,0}\makebox(0,0)[t]{\lineheight{0}\smash{\begin{tabular}[t]{c}1.83\end{tabular}}}}%
  \end{picture}%
\endgroup%

%% file: coilcurrent.pdf_tex
\begingroup%
  \makeatletter%
  \providecommand\color[2][]{%
    \errmessage{(Inkscape) Color is used for the text in Inkscape, but the package 'color.sty' is not loaded}%
    \renewcommand\color[2][]{}%
  }%
  \providecommand\transparent[1]{%
    \errmessage{(Inkscape) Transparency is used (non-zero) for the text in Inkscape, but the package 'transparent.sty' is not loaded}%
    \renewcommand\transparent[1]{}%
  }%
  \providecommand\rotatebox[2]{#2}%
  \ifx\svgwidth\undefined%
    \setlength{\unitlength}{295.99520946bp}%
    \ifx\svgscale\undefined%
      \relax%
    \else%
      \setlength{\unitlength}{\unitlength * \real{\svgscale}}%
    \fi%
  \else%
    \setlength{\unitlength}{\svgwidth}%
  \fi%
  \global\let\svgwidth\undefined%
  \global\let\svgscale\undefined%
  \makeatother%
  \begin{picture}(1,0.72642557)%
    \put(0,0){\includegraphics[width=\unitlength,page=1]{coilcurrent.pdf}}%
    \put(0.24977449,0.04391526){\color[rgb]{0,0,0}\makebox(0,0)[b]{\smash{-20}}}%
    \put(0.4128149,0.04391526){\color[rgb]{0,0,0}\makebox(0,0)[b]{\smash{0}}}%
    \put(0.5707959,0.04391526){\color[rgb]{0,0,0}\makebox(0,0)[b]{\smash{20}}}%
    \put(0.72891933,0.04391526){\color[rgb]{0,0,0}\makebox(0,0)[b]{\smash{40}}}%
    \put(0.88605463,0.04391526){\color[rgb]{0,0,0}\makebox(0,0)[b]{\smash{60}}}%
    \put(0.1181351,0.26598687){\color[rgb]{0,0,0}\makebox(0,0)[rb]{\smash{0.05}}}%
    \put(0.12148017,0.43064508){\color[rgb]{0,0,0}\makebox(0,0)[rb]{\smash{0.1}}}%
    \put(0.1181351,0.59375032){\color[rgb]{0,0,0}\makebox(0,0)[rb]{\smash{0.15}}}%
    \put(0.87881471,0.62171134){\color[rgb]{0,0,0}\makebox(0,0)[lb]{\smash{Coil 1}}}%
    \put(0.87881471,0.5782169){\color[rgb]{0,0,0}\makebox(0,0)[lb]{\smash{Coil 2}}}%
    \put(0.87881471,0.53316941){\color[rgb]{0,0,0}\makebox(0,0)[lb]{\smash{Coil 3}}}%
    \put(1.00204989,0.00430012){\color[rgb]{0,0,0}\makebox(0,0)[rb]{\smash{Time from extraction of $\overline{p}$ [ms]}}}%
    \put(0.02599818,0.672935){\color[rgb]{0,0,0}\rotatebox{90}{\makebox(0,0)[rb]{\smash{$B_z$ on axis [T]}}}}%
    \put(0.11945075,0.0990999){\color[rgb]{0,0,0}\makebox(0,0)[rb]{\smash{0}}}%
  \end{picture}%
\endgroup%

%% file: retardpot.pdf_tex
\begingroup%
  \makeatletter%
  \providecommand\color[2][]{%
    \errmessage{(Inkscape) Color is used for the text in Inkscape, but the package 'color.sty' is not loaded}%
    \renewcommand\color[2][]{}%
  }%
  \providecommand\transparent[1]{%
    \errmessage{(Inkscape) Transparency is used (non-zero) for the text in Inkscape, but the package 'transparent.sty' is not loaded}%
    \renewcommand\transparent[1]{}%
  }%
  \providecommand\rotatebox[2]{#2}%
  \ifx\svgwidth\undefined%
    \setlength{\unitlength}{326.40984242bp}%
    \ifx\svgscale\undefined%
      \relax%
    \else%
      \setlength{\unitlength}{\unitlength * \real{\svgscale}}%
    \fi%
  \else%
    \setlength{\unitlength}{\svgwidth}%
  \fi%
  \global\let\svgwidth\undefined%
  \global\let\svgscale\undefined%
  \makeatother%
  \begin{picture}(1,0.98036263)%
    \put(0,0){\includegraphics[width=\unitlength,page=1]{retardpot.pdf}}%
    \put(0.92549288,-0.16955805){\color[rgb]{0,0,0}\makebox(0,0)[b]{\smash{}}}%
    \put(-0.00211729,0.90901171){\color[rgb]{0,0,0}\makebox(0,0)[lb]{\smash{(a)}}}%
    \put(0.15414539,0.64601714){\color[rgb]{0,0,0}\makebox(0,0)[rb]{\smash{-5}}}%
    \put(0.98314109,0.95855551){\color[rgb]{0,0,0}\makebox(0,0)[rb]{\smash{{\small Retarding potential on axis $V_r$}}}}%
    \put(0.33416876,0.79677661){\color[rgb]{0,0,0}\makebox(0,0)[lb]{\smash{$\overline{p}\rightarrow$}}}%
    \put(1.00239381,0.49992552){\color[rgb]{0,0,0}\makebox(0,0)[rb]{\smash{{\small Distance from the center of the MUSASHI trap [mm]}}}}%
    \put(0.08594765,0.93583705){\color[rgb]{0,0,0}\rotatebox{90}{\makebox(0,0)[rb]{\smash{{\small Potential on axis [V]}}}}}%
    \put(0.34134085,0.53147136){\color[rgb]{0,0,0}\makebox(0,0)[b]{\smash{{\small 0}}}}%
    \put(0.5852231,0.53147136){\color[rgb]{0,0,0}\makebox(0,0)[b]{\smash{{\small 400}}}}%
    \put(0.82838634,0.53147136){\color[rgb]{0,0,0}\makebox(0,0)[b]{\smash{{\small 800}}}}%
    \put(0.1536275,0.90206379){\color[rgb]{0,0,0}\makebox(0,0)[rb]{\smash{0}}}%
    \put(0.0833778,0.42630932){\color[rgb]{0,0,0}\rotatebox{90}{\makebox(0,0)[rb]{\smash{{\small Potential on axis [V]}}}}}%
    \put(0.64414777,0.18673156){\color[rgb]{0,0,0}\makebox(0,0)[rb]{\smash{{\small Retarding potential on axis $V_r$}}}}%
    \put(0.6000171,0.13700706){\color[rgb]{0,0,0}\makebox(0,0)[rb]{\smash{{\small Confinement well}}}}%
    \put(0.18727032,0.27015872){\color[rgb]{0,0,0}\makebox(0,0)[lb]{\smash{{\small $\overline{p}\rightarrow$}}}}%
    \put(0.14470901,0.29518746){\color[rgb]{0,0,0}\makebox(0,0)[rb]{\smash{0}}}%
    \put(0.14412859,0.41289799){\color[rgb]{0,0,0}\makebox(0,0)[rb]{\smash{4}}}%
    \put(0.14391744,0.1809304){\color[rgb]{0,0,0}\makebox(0,0)[rb]{\smash{-4}}}%
    \put(0.14423981,0.06406836){\color[rgb]{0,0,0}\makebox(0,0)[rb]{\smash{-8}}}%
    \put(1.00210795,0.00502042){\color[rgb]{0,0,0}\makebox(0,0)[rb]{\smash{{\small Distance from the center of the MUSASHI trap [mm]}}}}%
    \put(-0.00220563,0.40779333){\color[rgb]{0,0,0}\makebox(0,0)[lb]{\smash{(b)}}}%
    \put(0.86601053,0.3534309){\color[rgb]{0,0,0}\makebox(0,0)[rb]{\smash{{\small Potential well at $\overline{p}$ injection}}}}%
    \put(0.22491404,0.03975436){\color[rgb]{0,0,0}\makebox(0,0)[b]{\smash{{\small 2300}}}}%
    \put(0.52886743,0.03975436){\color[rgb]{0,0,0}\makebox(0,0)[b]{\smash{{\small 2500}}}}%
    \put(0.83281843,0.03975436){\color[rgb]{0,0,0}\makebox(0,0)[b]{\smash{{\small 2700}}}}%
    \put(0.54598675,0.64957981){\color[rgb]{0,0,0}\makebox(0,0)[lb]{\smash{($V_f=1.5$ V)}}}%
  \end{picture}%
\endgroup%

%% file: energydist.pdf_tex
\begingroup%
  \makeatletter%
  \providecommand\color[2][]{%
    \errmessage{(Inkscape) Color is used for the text in Inkscape, but the package 'color.sty' is not loaded}%
    \renewcommand\color[2][]{}%
  }%
  \providecommand\transparent[1]{%
    \errmessage{(Inkscape) Transparency is used (non-zero) for the text in Inkscape, but the package 'transparent.sty' is not loaded}%
    \renewcommand\transparent[1]{}%
  }%
  \providecommand\rotatebox[2]{#2}%
  \ifx\svgwidth\undefined%
    \setlength{\unitlength}{386.33703998bp}%
    \ifx\svgscale\undefined%
      \relax%
    \else%
      \setlength{\unitlength}{\unitlength * \real{\svgscale}}%
    \fi%
  \else%
    \setlength{\unitlength}{\svgwidth}%
  \fi%
  \global\let\svgwidth\undefined%
  \global\let\svgscale\undefined%
  \makeatother%
  \begin{picture}(1,1.0353653)%
    \put(0,0){\includegraphics[width=\unitlength,page=1]{energydist.pdf}}%
    \put(-0.00193805,1.01870245){\color[rgb]{0,0,0}\makebox(0,0)[lb]{\smash{(a)}}}%
    \put(0.02015798,1.00557314){\color[rgb]{0,0,0}\rotatebox{90}{\makebox(0,0)[rb]{\smash{{\small Normalized annihilations [$\mu$C]}}}}}%
    \put(-0.0019383,0.49586348){\color[rgb]{0,0,0}\makebox(0,0)[lb]{\smash{(b)}}}%
    \put(0.02015798,0.47068521){\color[rgb]{0,0,0}\rotatebox{90}{\makebox(0,0)[rb]{\smash{{\small Normalized annihilation count}}}}}%
    \put(0.97798186,0.00374036){\color[rgb]{0,0,0}\makebox(0,0)[rb]{\smash{$V_r$ [V]}}}%
    \put(0.17876113,0.03215579){\color[rgb]{0,0,0}\makebox(0,0)[b]{\smash{{\small $\mu-10$}}}}%
    \put(0.3597131,0.03215579){\color[rgb]{0,0,0}\makebox(0,0)[b]{\smash{{\small $\mu-5$}}}}%
    \put(0.72161681,0.03215579){\color[rgb]{0,0,0}\makebox(0,0)[b]{\smash{{\small $\mu+5$}}}}%
    \put(0.90256873,0.03215579){\color[rgb]{0,0,0}\makebox(0,0)[b]{\smash{{\small $\mu+10$}}}}%
    \put(0.54066491,0.03215579){\color[rgb]{0,0,0}\makebox(0,0)[b]{\smash{{\small $\mu$}}}}%
    \put(0.97798186,0.5386283){\color[rgb]{0,0,0}\makebox(0,0)[rb]{\smash{$V_r$ [V]}}}%
    \put(0.17876126,0.56704347){\color[rgb]{0,0,0}\makebox(0,0)[b]{\smash{{\small $\mu-10$}}}}%
    \put(0.35971323,0.56704347){\color[rgb]{0,0,0}\makebox(0,0)[b]{\smash{{\small $\mu-5$}}}}%
    \put(0.72161675,0.56704347){\color[rgb]{0,0,0}\makebox(0,0)[b]{\smash{{\small $\mu+5$}}}}%
    \put(0.90256873,0.56704347){\color[rgb]{0,0,0}\makebox(0,0)[b]{\smash{{\small $\mu+10$}}}}%
    \put(0.54066491,0.56704347){\color[rgb]{0,0,0}\makebox(0,0)[b]{\smash{{\small $\mu$}}}}%
    \put(0.66687359,0.9554662){\color[rgb]{0,0,0}\makebox(0,0)[lb]{\smash{{\small $\blacktriangledown$ $V_f=1.5$~V}}}}%
    \put(0.66687359,0.91246903){\color[rgb]{0,0,0}\makebox(0,0)[lb]{\smash{{\Large $\bullet$} {\small $V_f=20$~V}}}}%
    \put(0.07092622,0.61585635){\color[rgb]{0,0,0}\makebox(0,0)[lb]{\smash{0}}}%
    \put(0.05664094,0.66664685){\color[rgb]{0,0,0}\makebox(0,0)[lb]{\smash{0.2}}}%
    \put(0.05664094,0.71902423){\color[rgb]{0,0,0}\makebox(0,0)[lb]{\smash{0.4}}}%
    \put(0.05664094,0.76981485){\color[rgb]{0,0,0}\makebox(0,0)[lb]{\smash{0.6}}}%
    \put(0.05664094,0.82060547){\color[rgb]{0,0,0}\makebox(0,0)[lb]{\smash{0.8}}}%
    \put(0.07410075,0.8698092){\color[rgb]{0,0,0}\makebox(0,0)[lb]{\smash{1}}}%
    \put(0.05664094,0.92218646){\color[rgb]{0,0,0}\makebox(0,0)[lb]{\smash{1.2}}}%
    \put(0.05664094,0.97297695){\color[rgb]{0,0,0}\makebox(0,0)[lb]{\smash{1.4}}}%
    \put(0.07092622,0.08096752){\color[rgb]{0,0,0}\makebox(0,0)[lb]{\smash{0}}}%
    \put(0.0471175,0.12699644){\color[rgb]{0,0,0}\makebox(0,0)[lb]{\smash{0.05}}}%
    \put(0.05981533,0.17461276){\color[rgb]{0,0,0}\makebox(0,0)[lb]{\smash{0.1}}}%
    \put(0.0471175,0.22064144){\color[rgb]{0,0,0}\makebox(0,0)[lb]{\smash{0.15}}}%
    \put(0.05664094,0.26825839){\color[rgb]{0,0,0}\makebox(0,0)[lb]{\smash{0.2}}}%
    \put(0.0471175,0.31428668){\color[rgb]{0,0,0}\makebox(0,0)[lb]{\smash{0.25}}}%
    \put(0.05664094,0.36031497){\color[rgb]{0,0,0}\makebox(0,0)[lb]{\smash{0.3}}}%
    \put(0.0471175,0.40793193){\color[rgb]{0,0,0}\makebox(0,0)[lb]{\smash{0.35}}}%
    \put(0.05664094,0.45396022){\color[rgb]{0,0,0}\makebox(0,0)[lb]{\smash{0.4}}}%
    \put(0.66687359,0.41795246){\color[rgb]{0,0,0}\makebox(0,0)[lb]{\smash{{\small $\blacktriangledown$ $V_f=1.5$~V}}}}%
    \put(0.66687359,0.37495529){\color[rgb]{0,0,0}\makebox(0,0)[lb]{\smash{{\Large $\bullet$} {\small $V_f=20$~V}}}}%
    \put(0.9592453,0.65832005){\color[rgb]{0,0,0}\makebox(0,0)[rb]{\smash{MUSASHI}}}%
    \put(0.95866139,0.12622199){\color[rgb]{0,0,0}\makebox(0,0)[rb]{\smash{CUSP}}}%
  \end{picture}%
\endgroup%

%% file: potfi.pdf_tex
\begingroup%
  \makeatletter%
  \providecommand\color[2][]{%
    \errmessage{(Inkscape) Color is used for the text in Inkscape, but the package 'color.sty' is not loaded}%
    \renewcommand\color[2][]{}%
  }%
  \providecommand\transparent[1]{%
    \errmessage{(Inkscape) Transparency is used (non-zero) for the text in Inkscape, but the package 'transparent.sty' is not loaded}%
    \renewcommand\transparent[1]{}%
  }%
  \providecommand\rotatebox[2]{#2}%
  \ifx\svgwidth\undefined%
    \setlength{\unitlength}{267.38203154bp}%
    \ifx\svgscale\undefined%
      \relax%
    \else%
      \setlength{\unitlength}{\unitlength * \real{\svgscale}}%
    \fi%
  \else%
    \setlength{\unitlength}{\svgwidth}%
  \fi%
  \global\let\svgwidth\undefined%
  \global\let\svgscale\undefined%
  \makeatother%
  \begin{picture}(1,0.6522501)%
    \put(0,0){\includegraphics[width=\unitlength,page=1]{potfi.pdf}}%
    \put(0.03796941,0.65429581){\color[rgb]{0,0,0}\rotatebox{90}{\makebox(0,0)[rb]{\smash{{\footnotesize Potential on axis [V]}}}}}%
    \put(0.96780384,0.00487182){\color[rgb]{0,0,0}\makebox(0,0)[rb]{\smash{{\footnotesize Distance from the center of the MUSASHI trap [mm]}}}}%
    \put(0.50867204,0.04706213){\color[rgb]{0,0,0}\makebox(0,0)[b]{\smash{{\footnotesize 2800}}}}%
    \put(0.91912754,0.04873295){\color[rgb]{0,0,0}\makebox(0,0)[b]{\smash{{\footnotesize 3000}}}}%
    \put(0.45440042,0.30468706){\color[rgb]{0,0,0}\makebox(0,0)[lb]{\smash{{\small $e^+$}}}}%
    \put(0.1576388,0.2620014){\color[rgb]{1,0,1}\makebox(0,0)[lb]{\smash{{\small $\overline{p}$ injection energy}}}}%
    \put(0.75986485,0.59302711){\color[rgb]{0,0,0}\makebox(0,0)[rb]{\smash{{\footnotesize Steep well for field ionization (FI)}}}}%
    \put(0.11655342,0.61789366){\color[rgb]{0,0,0}\makebox(0,0)[rb]{\smash{{\footnotesize 200}}}}%
    \put(0.11655342,0.4866906){\color[rgb]{0,0,0}\makebox(0,0)[rb]{\smash{{\footnotesize 100}}}}%
    \put(0.11655342,0.35548115){\color[rgb]{0,0,0}\makebox(0,0)[rb]{\smash{{\footnotesize 0}}}}%
    \put(0.11655342,0.22427169){\color[rgb]{0,0,0}\makebox(0,0)[rb]{\smash{{\footnotesize -100}}}}%
    \put(0.11655342,0.09305585){\color[rgb]{0,0,0}\makebox(0,0)[rb]{\smash{{\footnotesize -200}}}}%
    \put(0.37191028,0.45108275){\color[rgb]{0,0,0}\makebox(0,0)[b]{\smash{{\footnotesize Nested well}}}}%
  \end{picture}%
\endgroup%

%% file: hbarint.pdf_tex
\begingroup%
  \makeatletter%
  \providecommand\color[2][]{%
    \errmessage{(Inkscape) Color is used for the text in Inkscape, but the package 'color.sty' is not loaded}%
    \renewcommand\color[2][]{}%
  }%
  \providecommand\transparent[1]{%
    \errmessage{(Inkscape) Transparency is used (non-zero) for the text in Inkscape, but the package 'transparent.sty' is not loaded}%
    \renewcommand\transparent[1]{}%
  }%
  \providecommand\rotatebox[2]{#2}%
  \ifx\svgwidth\undefined%
    \setlength{\unitlength}{340.04711962bp}%
    \ifx\svgscale\undefined%
      \relax%
    \else%
      \setlength{\unitlength}{\unitlength * \real{\svgscale}}%
    \fi%
  \else%
    \setlength{\unitlength}{\svgwidth}%
  \fi%
  \global\let\svgwidth\undefined%
  \global\let\svgscale\undefined%
  \makeatother%
  \begin{picture}(1,0.5881538)%
    \put(0.0164121,0.71356242){\color[rgb]{0,0,0}\rotatebox{90}{\makebox(0,0)[rb]{\smash{}}}}%
    \put(0,0){\includegraphics[width=\unitlength,page=1]{hbarint.pdf}}%
    \put(0.99326674,0.55236726){\color[rgb]{0,0,0}\makebox(0,0)[rb]{\smash{{\Large $\bullet$} {\footnotesize Reduced energy spread in 2016 ($0.23\pm0.02$~eV)}}}}%
    \put(0.99326674,0.51837773){\color[rgb]{0,0,0}\makebox(0,0)[rb]{\smash{{\footnotesize $\blacksquare$ Previous condition in 2012 (15~eV)}}}}%
    \put(0.02788745,0.58871839){\color[rgb]{0,0,0}\rotatebox{90}{\makebox(0,0)[rb]{\smash{{\footnotesize Field ionized $\overline{\rm H}$ atoms}}}}}%
    \put(0.08495665,0.47541983){\color[rgb]{0,0,0}\makebox(0,0)[rb]{\smash{{\footnotesize 300}}}}%
    \put(0.08495665,0.33555587){\color[rgb]{0,0,0}\makebox(0,0)[rb]{\smash{{\footnotesize 200}}}}%
    \put(0.08495665,0.19569665){\color[rgb]{0,0,0}\makebox(0,0)[rb]{\smash{{\footnotesize 100}}}}%
    \put(0.08495665,0.05583255){\color[rgb]{0,0,0}\makebox(0,0)[rb]{\smash{{\footnotesize 0}}}}%
    \put(0.11272378,0.03735156){\color[rgb]{0,0,0}\makebox(0,0)[b]{\smash{{\footnotesize 0}}}}%
    \put(0.35004469,0.03734624){\color[rgb]{0,0,0}\makebox(0,0)[b]{\smash{{\footnotesize 20}}}}%
    \put(0.58792,0.03735156){\color[rgb]{0,0,0}\makebox(0,0)[b]{\smash{{\footnotesize 40}}}}%
    \put(0.82505669,0.03735156){\color[rgb]{0,0,0}\makebox(0,0)[b]{\smash{{\footnotesize 60}}}}%
    \put(1.00177358,0.00948962){\color[rgb]{0,0,0}\makebox(0,0)[rb]{\smash{{\footnotesize Elapsed time from injection of $\overline{p}$ beam [s]}}}}%
  \end{picture}%
\endgroup%